\documentclass{emulateapj}
\usepackage{natbib,graphicx}

\def\bnabla{\mbox{\boldmath $\nabla$}}

\def\refnew#1{(\ref{#1})}
\def\be{\begin{equation}}
\def\ee{\end{equation}}

\def\s{\, \rm s}
\def\K{\, \rm K}

\def\s{\, \rm s}
\def\km{\, \rm km}
\def\cm{\, \rm cm}

\def\dyne{\, \rm dyne}

\def\g{\rm g}

\def\bov{{\bf v}}
\def\bE{{\bf E}}
\def\bB{{\bf B}}
\def\bj{{\bf j}}
\def\bM{{\bf M}}
\def\ber{{\bf e}_r}
\def\bet{{\bf e}_\theta}
\def\br{{\bf r}}
\def\bnabla{{\bf \nabla}}
\def\bof{{\bf f}}

\defcitealias{Batygin-ohmic}{B12}
\usepackage{ulem}  
\usepackage{color}

\newcommand{\y}{\color{black}}

\newcommand{\beqn}{\begin{eqnarray}}
\newcommand{\eeqn}{\end{eqnarray}}
\newcommand{\bld}[1]{\mbox{\boldmath$#1$\unboldmath}}

\begin{document} 
		
\title{\mbox{{Ohmic Heating Suspends, not Reverses, the Cooling
      Contraction of Hot Jupiters} }}

\author{Yanqin Wu\altaffilmark{1}, 
Yoram Lithwick\altaffilmark{2}}
\altaffiltext{1}{Department of Astronomy \& Astrophysics, University
  of Toronto, Toronto, ON M5S 3H4, Canada} 
\altaffiltext{2}{Department of Physics \& Astronomy, Northwestern
  University, Evanston, IL 60208 {\y \& Center for Interdisciplinary Exploration and Research in Astrophysics (CIERA)}}

\begin{abstract}
  We study the radius evolution of close-in extra-solar jupiters under
  Ohmic heating, a mechanism that was recently proposed to explain the
  large observed sizes of many of these planets.  Planets are born
  with high entropy and they subsequently cool and contract. We focus
  on two cases: first, that ohmic heating commences when the planet is
  hot (high entropy); and second, that it commences after the planet
  has cooled. In the former case, we use analytical scaling and
  numerical experiments to confirm that Ohmic heating is capable of
  suspending the cooling as long as a few percent of the stellar
  irradiation is converted into Ohmic heating, and the planet has a
  surface wind that extends to pressures of $\sim 10$ bar or deeper.
  For these parameters, the radii at which cooling is stalled are
  consistent with (or larger than) the observed radii of most
  planets. The only two exceptions are WASP-17b and HAT-P-32b.  In
  contrast to the high entropy case, we show that Ohmic heating cannot
  significantly re-inflate planets after they have already cooled.
  {This leads us to suggest} that the diversity of radii observed
  in hot jupiters may be partially explained {by the different
    epochs at which they are migrated to their current locations.}
\end{abstract}

\setcounter{equation}{0}


\section{Observed Radii of Hot Jupiters}
Since the discovery of the first transiting hot jupiter
\citep{Charbonneau}, the radii of these objects have remained
puzzling. While cooling theory predicts that these planets, regardless
of mass, should contract to $\sim 1 R_J$ (Jupiter radius) at the age
of a few Gyrs, the observed radii for hot jupiters range from $0.8
R_J$ to $2 R_J$. Factors such as stellar irradiation, a varying core
mass, and atmospheric metal content can at best cause $\sim 10\%
-20\%$ variations in the radius \citep[see, e.g.][]{Burrows}.


There is a trend that the observed radii rise with the amount of
stellar irradiation \citep[e.g.,][also see data in
Fig. \ref{fig:ohmic-radius}] {Enoch,Laughlin11} though the most
inflated planet discovered to date \citep[WASP-17b,][]{wasp17b} is not
the hottest one.

Lacking a mechanism of energy generation, isolated Jovian planets
inexorably contract as they cool. The inflated hot jupiters look much
younger than their real ages suggest. Some mechanism is at work either
to stall their aging or to rejuvenate them from old age \citep[{\y see
  proposals by,
  e.g.,}][]{2001ApJ...548..466B,Chabrier,2010ApJ...721.1113Y,Batygin-ohmic0}.
The rejuvenation may be particularly relevant for hot jupiters, as
they are believed not to have formed in situ, but to have migrated
inward after they formed (perhaps well after).  We consider in detail
the proposal of \citet{Batygin-ohmic0} {where a surface wind
  \citep[see review by][]{2010exop.book..471S} blowing across the
  planetary magnetic field acts as a battery that sends current to the
  interior and gives rise to Ohmic heating in the deeper layers.  {\y
    A number of other studies have also discussed Ohmic heating in the
    atmospheres of hot jupiters \citep{Perna,Heng,menou11}. }  {\y Our
    study here is similar in spirit to that of \citet{Batygin-ohmic}
    (hereafter B12), where they incorporat Ohmic heating (both in the
    atmosphere and in deeper layers) into calculations of planet
    thermal evolutions. They find that Ohmic heating can explain the
    inflated sizes of hot jupiters, and that low mass hot jupiters can
    be inflated to become Roche lobe overflowing.} In this work, we
  {\y also} find that Ohmic heating appears effective in stalling the
  cooling contraction, largely confirming results of
  \citetalias{Batygin-ohmic}, but {\y we find that} it is incapable of
  re-inflating the planets {\y after they have contracted.}

  Our work builds on that of \citet{Batygin-ohmic0} but we focus on
  the thermodynamical aspects; we elucidate the condition for Ohmic
  heating to be effective; and we investigate the interior structure
  of planets as affected by Ohmic heating. {\y We are particularly
    interested in the question of planet rejuvenation. Our work shows
    that Ohmic heating, being a largely surface phenomenon, is
    ineffective at re-inflating planets, as this requires entropy
    deposition in the very center of planets.}

\section{Stalling Contraction}
\label{sec:thermal}

\subsection{General Criterion for Stalling Contraction}

We consider the stalling of contraction in an irradiated planet.  An
elucidating work in this regard is
\citet{ArrasBildsten}. Our discussion here is similar in spirit to
theirs. 

The interior of the planet is largely convective due to the high
opacity there. The resulting temperature profile is adiabatic, with an
internal entropy $S$. The layer below the photosphere, however, is
nearly isothermal with $T \approx T_{\rm eq}$, as a result of the
strong stellar irradiation.  We simplify the planet into a two-zone
model, an upper isothermal envelope and an isentropic core (see
Fig. \ref{fig:noohmic-special-T}).\footnote{\y This simplification may
  break down when, e.g., due to Ohmic heating, another adiabatic zone
  appears near the surface. We ignore this complication here for
  clarity of exposition.} We denote the depth of the transition point
as $z_{\rm tr}$. This depth determines the cooling rate of the planet.

\begin{figure}[t]
\begin{center}
\includegraphics[width=0.50\textwidth,angle=0,
trim=0 20 170 230,clip=true]{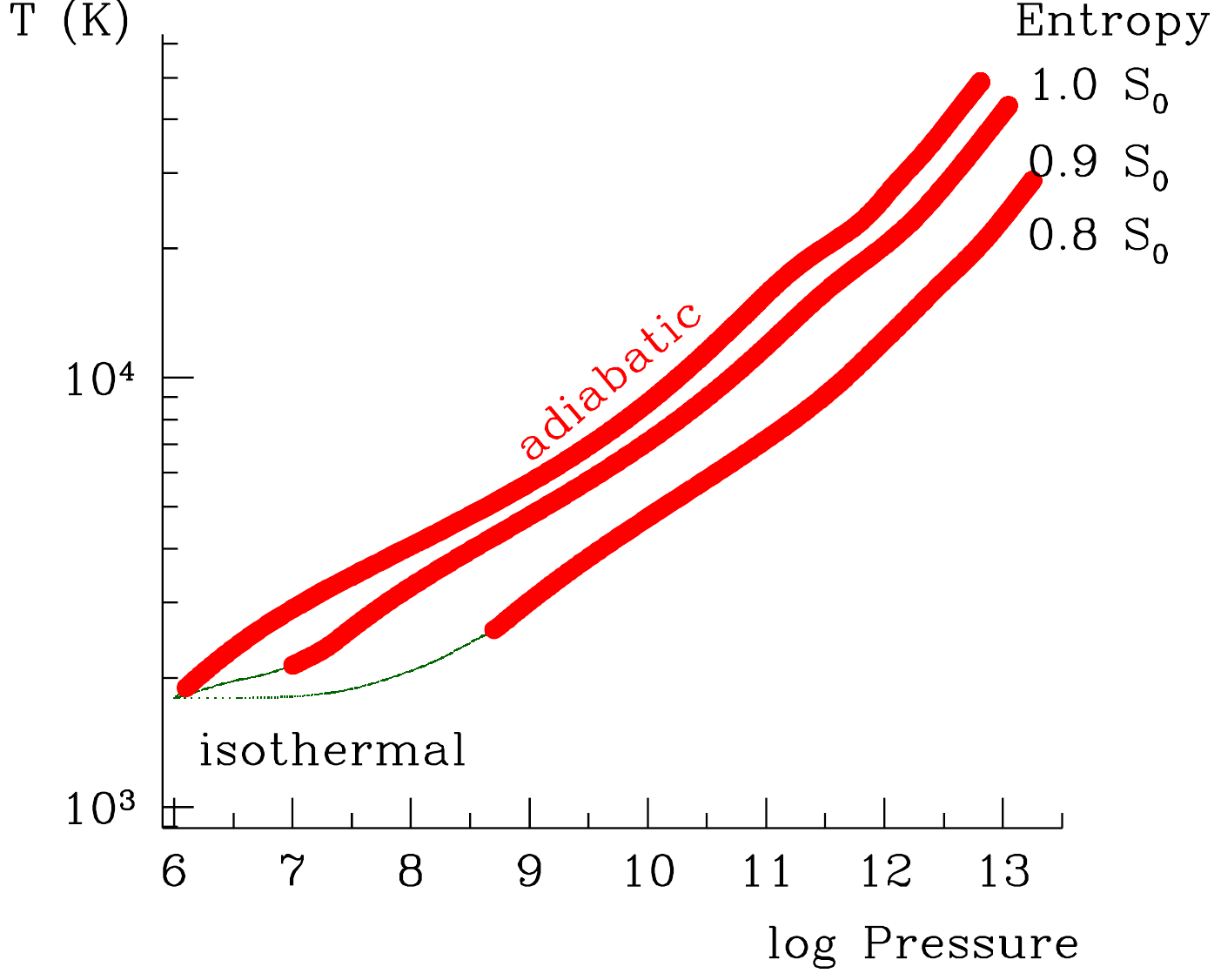}
\caption{Temperature as a function of logarithmic pressure (in cgs
  units) inside an irradiated planet, as it cools with time. Values
  for the equilibrium temperature and the planet mass are chosen to
  resemble those of TrEs4-b: $T_{\rm eq} = 1788 \K$ and $M_p = 0.91
  M_J$. Shown here are the three temperature profiles when internal
  entropy (in the isentropic core) takes values of $S=1 S_0$, $0.9
  S_0$ and $0.8 S_0$, with $S_0$ being value of the surface entropy.
  There is a one-to-one correspondence between $S$ and the planet
  radius (which progresses from $1.9$ to $1.5$ to $1.2
  R_J$). Moreover, since the convection zone (marked in red) recedes
  as the planet cools, there is also a one-to-one correspondence
  between $S$ and the self luminosity (which decreases from $10^{-4}$
  to $10^{-6}$ to $10^{-7} L_\odot$).  These models are obtained using
  our thermal equilibrium code (\S \ref{sec:nr}).  }
\label{fig:noohmic-special-T}
\end{center}
\end{figure}

As we now demonstrate, the value of the internal entropy ($S$)
determines both the planet's radius ($R$) and its cooling luminosity
($L$).  The first connection is straightforward -- a higher $S$
corresponds to a higher internal temperature and therefore a larger
scale height. The size of the planet is mostly determined by its
bottom-most scale height. So in a given planet, $R$ is a monotonic
function of $S$.

To establish the connection between $S$ and $L$, it is useful to
recognize that the planet's cooling is controlled exclusively by the
optical depth and the temperature gradient at the transition point,
$z_{\rm tr}$ \citep{ArrasBildsten}.  Radiative diffusion allows an
energy leakage that scales with the temperature gradient as,
\begin{equation}
F_{\rm rad}\,\,\,  =\,\,\, {{ a c}\over{3 \kappa \rho}} {{dT^4}\over{dz}} 
= \,\,\, {{4 \sigma_B}\over 3}\, {{d T^4}\over{d\tau}}, 
\label{eq:Frad}
\end{equation}
where $\tau = \int_0^z \kappa\rho dz^\prime$ is the optical depth
measured from the surface, and $\sigma_B$ the Stefan-Boltzman
constant.  Below $z_{\rm tr}$, the temperature gradient is adiabatic
and the optical depth rises quickly; above $z_{\rm tr}$, the
temperature is nearly isothermal, $T_{\rm tr} \sim T_{\rm eq}$.  The
radiative flux of the entire planet is therefore set by the optical
depth and the (adiabatic) temperature gradient at $z_{\rm tr}$, and is
roughly $\sim \sigma_B T_{\rm eq}^4/\tau_{\rm tr}$.  
And lastly, since a 
 planet { with a cooler interior} has a deeper isothermal layer
(larger $z_{\rm tr}$, see Fig. \ref{fig:noohmic-special-T}), it will
therefore have a lower cooling flux.

Hence measuring the radius directly probes the internal entropy, which
in turn, informs us how fast the planet is losing energy. An inflated
planet has a higher entropy, and therefore has to lose heat
faster. The only way to stall contraction is to supply this cooling
luminosity, not from the internal energy of the planet, but from some
other means (e.g., Ohmic heating, wave energy, etc.).  Moreover, this
extra energy source has to be able to replace the cooling flux at
depth $z \geq z_{\rm tr}$.

{\y In our calculations, we frequently find that a new convection zone
  develops in the wind zone, where Ohmic dissipation is at its
  greatest power. This has the effect of pushing $z_{\rm tr}$ to a
  deeper depth, for the same interior adiabat. But it will not change
  the above discussion.}



\subsection{Constraints on Ohmic Heating}

In this paper, we focus on the Ohmic heating mechanism
\citep{Batygin-ohmic0}, which is one of the best laid-out mechanisms
for halting contraction. In this model, if the planet has a large
scale magnetic field that is anchored in the deep interior, and if its
poorly conducting surface layer is differentially rotating relative to
the interior, the current generated by pulling the field line in the
surface wind layer returns through the interior, producing Ohmic
heating in the interior.\footnote{\y We ignore current leakage from
  the top of the atmosphere, important for very hot atmospheres that
  sustain a significant ionization fraction \citep{Perna}.}

As established above, the relevant part of heating for stalling the
planet's contraction is the part that is deposited below $z_{\rm tr}$.
The general criterion for stalling heating then translates to
\begin{equation}
  \int_R^{z_{\rm tr}} \left.{{dQ}\over{dt}}\right|_{\rm Ohmic} dz \approx  F_{rad}|_{z_{\rm tr}}
  \approx {{\sigma_B T_{\rm eq}^4}\over{\tau_{\rm tr}}}
\label{eq:translate}
\end{equation}
We express the planet's equilibrium temperature as $\sigma_B$ $T_{\rm
  eq}^4 \sim L_*/ 16\pi a^2$ with $L_*$ the stellar luminosity and $a$
the stellar-planet separation.  If an abnormal opacity source is
present and causes a strong greenhouse effect, the value of $T_{\rm
  eq}$ will be higher than adopted here.

The volumetric rate for Ohmic heating $dQ/dt = J^2/\sigma$, where $J$
is the electric current density and $\sigma$ is the electric
conductivity.  There are two characteristics of the profiles of Ohmic
heating that pertain to this discussion, both demonstrated in Appendix
\ref{sec:radial} via an analytical model.  First, as wind in the
shallow layer pulls the poloidal magnetic field forward, a perturbed
toroidal magnetic field, $b_\phi$, and a largely meridional current,
${\bf J} \approx J_\theta {\bf e_\theta}$, are created.  Since the
current is divergence-free, by geometry the radial current $J_r$ in
this thin layer is roughly $\delta =z_{\rm wind}/R \ll 1$ smaller than
the meridional current, where $z_{\rm wind}$ is the depth of the wind
layer (eq. \ref{eq:outer}). The radial current flows continuously
across the bottom of the wind zone, generating an interior radial
current $J_r$ of a comparable magnitude (eq. \ref{eq:middle}). The
interior $J_\theta$, however, is comparable to the local $J_r$ in this
geometrically thicker zone (eq. \ref{eq:inner}).  As a result the
current density drops at the interface from $J \approx J_\theta \sim
J_r\, R/z_{\rm wind}$ above it to $J_r$ below.  The local heating
rate, $J^2/\sigma$, drops by a large factor of order $(R/z_{\rm
  wind})^2$ across the bottom of the wind zone. The bulk of the
integrated Ohmic heating is deposited inside the wind layer, with only
a fraction of order $(z_{\rm wind}/R)$ deposited below it
(Fig. \ref{fig:ohmic-special-compare-3}). {\y Our wind layer has a
  constant wind speed, but this discussion applies when, e.g. the wind
  speed varies sinusoidally with depth as in \citet{Batygin-ohmic0},
  or decreases linearly with logarithmic pressure.  }

Second, since the interior current density $J \approx J_r$ is
roughly constant below the wind layer, the interior volumetric heating
rate decays as $1/\sigma$.  Since $\sigma$ rises rapidly with depth,
most of the interior heating occurs right below the wind zone.

Following \citet{Batygin-ohmic0}, we adopt a heating efficiency
parameter $\epsilon$ that relates the total Ohmic heating (dominated
by that in the wind layer) to the stellar insolation impinging on the
planet,
\begin{equation}
\int_R^{0} \, {{dQ}\over{dt}} dz \approx 
{{J_{\rm wind}^2}\over{\sigma_{\rm wind}}} z_{\rm wind}
\approx \epsilon {{L_*}\over{16 \pi a^2}}\, ,
\end{equation}
where $J_{\rm wind}$ and $\sigma_{\rm wind}$ are evaluated in the wind
layer.  Now consider the part of Ohmic heating that occurs below
$z_{\rm tr}$. We express the current there as $J_{\rm tr} \approx
J_{\rm wind} (z_{\rm wind}/R) $ and find
\begin{equation}
\int_R^{z_{\rm tr}}\, {{dQ}\over{dt}} dz
 \approx  {{J_{\rm tr}^2}\over{\sigma_{\rm tr}}} z_{\rm tr}
\sim \epsilon {{L_*}\over{16 \pi a^2}} \left({{z_{\rm wind}}\over R}\right)^2
{{z_{\rm tr}}\over{z_{\rm wind}}} 
\left({{\sigma_{\rm wind}}\over{\sigma_{\rm tr}}}\right).
\end{equation}
Combining eq. \refnew{eq:translate} and the definition for $T_{\rm
  eq}$, we obtain a constraint on $\epsilon$ if Ohmic heating is to be
capable of stalling contraction,
\begin{equation}
\epsilon \geq
{1\over {\tau_{\rm tr} }}
\left({R\over{z_{\rm wind}}}\right)^2
{{z_{\rm wind}}\over{z_{\rm tr}}} 
\left({{\sigma_{\rm tr}}\over{\sigma_{\rm wind}}}\right).
\label{eq:epscon}
\end{equation}
A larger $\epsilon$ is required to sustain a larger (higher entropy)
planet, primarily because a higher entropy planet means a shallower
$z_{\rm tr}$ and a markedly smaller $\tau_{\rm tr}$. If the wind
penetrates only to a shallower depth $z_{\rm wind}$, a higher
$\epsilon$ is required to sustain the interior cooling.  Note that
this expression does not contain explicit dependence on the distance
to the star. However, the conductivity in the wind zone ($\sigma_{\rm
  wind}$) drops with a smaller $T_{\rm eq}$, in which case it becomes
harder to sustain the cooling using Ohmic heating.

\subsection{Numerical Results}
\label{sec:nr}

\begin{figure}
\begin{center}
\includegraphics[width=0.50\textwidth,angle=0,
trim=0 0 0 20,clip=true]{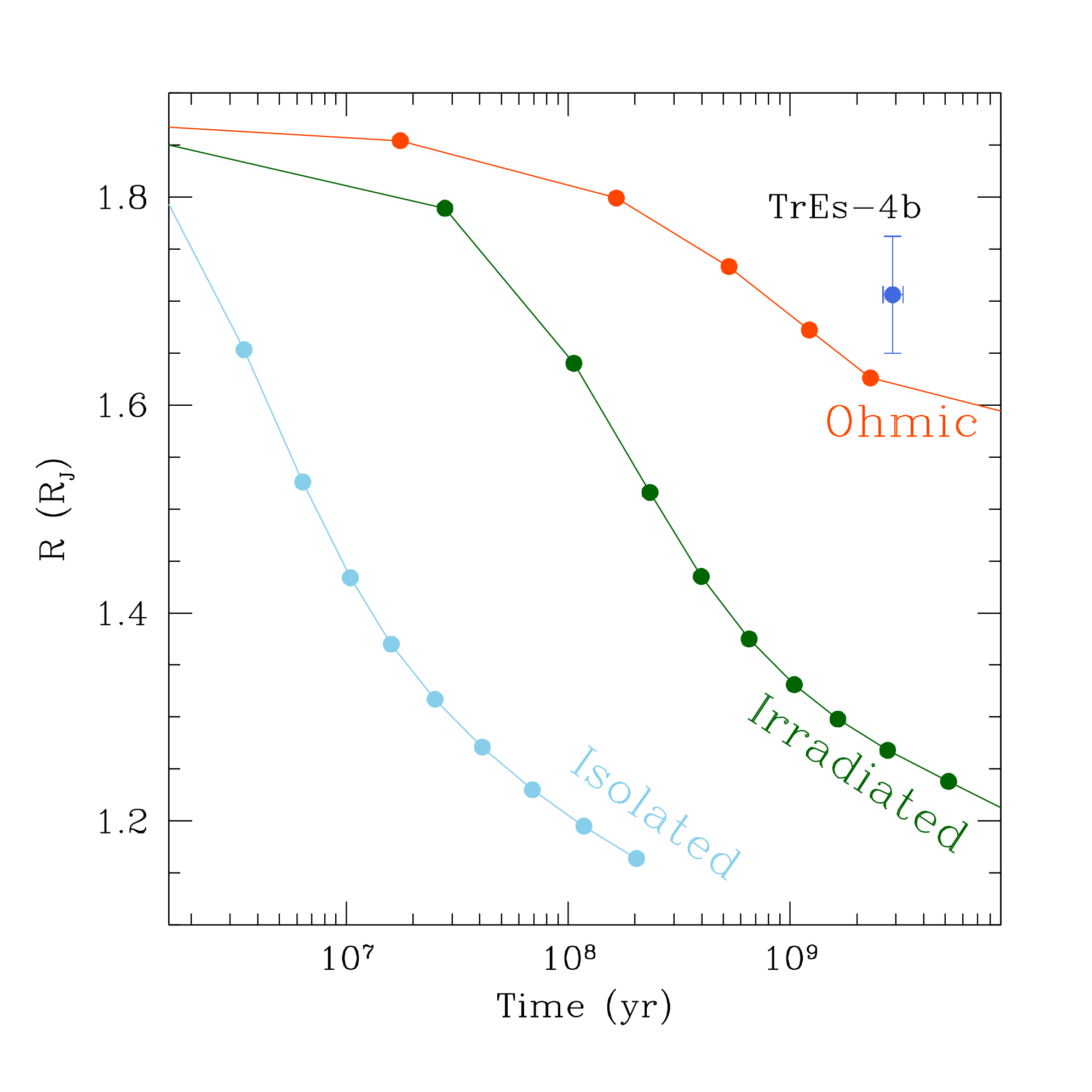}
\caption{Cooling curves for TrEs4-b if it is in isolation (lowest
  curve), irradiated (middle curve) or receives both irradiation and
  Ohmic heating (upper curve). We adopt $\epsilon = 3\%$ and $z_{\rm
    wind} = 2.5\times 10^8 \cm$. The Ohmic mechanism stalls the
  cooling by replacing most of the cooling luminosity at the top of
  the convection zone with ohmic heating. The data point indicates the
  measurement for TrEs4-b.}
\label{fig:ohmic-noohmic-special}
\end{center}
\end{figure}

\begin{figure}
\begin{center}
\includegraphics[width=0.50\textwidth,angle=0,
trim=30 120 30 120,clip=true]{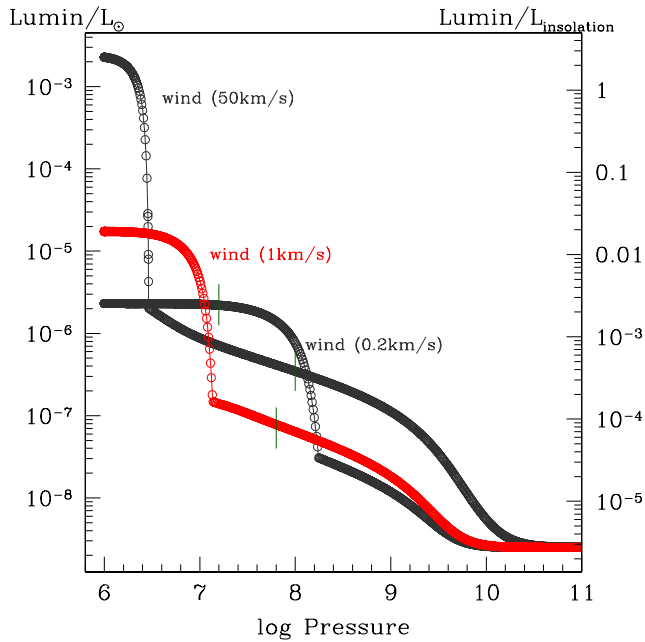}
\caption{The radial profile of Ohmic heating, plotted as integrated
  luminosity (relative to solar luminosity on the left axis, and
  relative to insolation luminosity on the right axis) against
  logarithmic pressure (in cgs unit), {\y using parameters similar to
    TrEs4-b, and} when the planet has a radius of $1.6 R_J$. The
  features in the profile, including the jump at $z_{\rm wind}$, are
  explained by our toy model in Appendix \ref{sec:radial}.  The middle
  curve is our standard case with $z_{\rm wind} = 2.5\times 10^8 \cm$
  (at a pressure of $\sim 10$ bar $=10^7 \dyne/\cm^2$), while the
  other two are $z_{\rm wind} = 8\times 10^7 \cm$ ($\sim 3$ bar), and
  $5\times 10^8\cm$ ($\sim 100$ bar), respectively.  The large size of
  the planet requires a high internal entropy and a high luminosity
  at the top of the convection zone, marked here by the short vertical bars.  
  This  luminosity can be  largely supplied
  by Ohmic heating if the wind is sufficiently deep and if the efficiency
is sufficiently high (eq. \refnew{eq:epscon}). The required  Ohmic efficiency is
 $3\%$  for our standard case, and $0.3\%$ and $200\%$ for the deeper
and shallower cases, respectively.
}
\label{fig:ohmic-special-compare-3}
\end{center}
\end{figure}

\begin{figure}
\begin{center}
\includegraphics[width=0.50\textwidth,angle=0,
trim=20 160 30 120,clip=true]{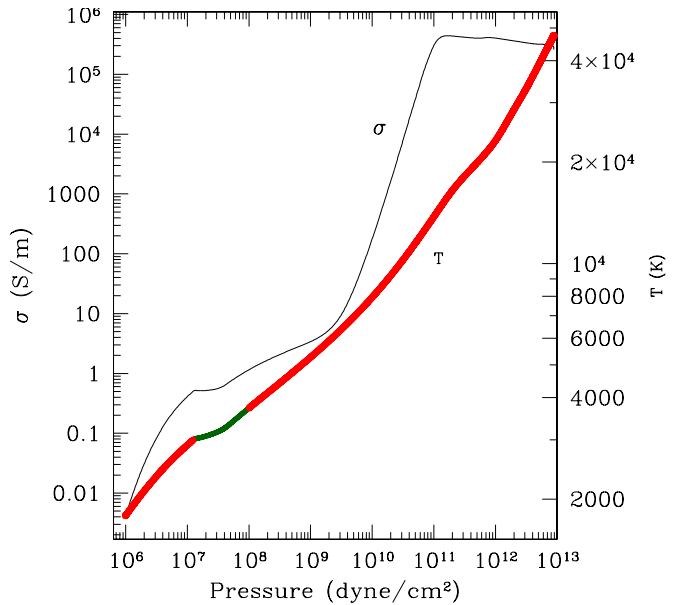}
\caption{{\y Electrical conductivity and gas temperature in a thermal
    equilibrium model for TrEs4-b, when subject to irradiation and
    Ohmic heating (according to the red 1 km/s curve in
    Fig. \ref{fig:ohmic-special-compare-3}). Near the surface,
    electrical conductivity is contributed by thermally ionized
    electrons from metals. Towards the interior, the contribution from
    hydrogen dominates.  The temperature is shown in red when the
    region is convective and green when stably stratified. The entire
    wind zone is convective.} }
\label{fig:ohmic-sigma-profile}
\end{center}
\end{figure}

\begin{figure*}
\begin{center}
\includegraphics[width=0.80\textwidth,angle=0,
trim=35 120 35 150,clip=true]{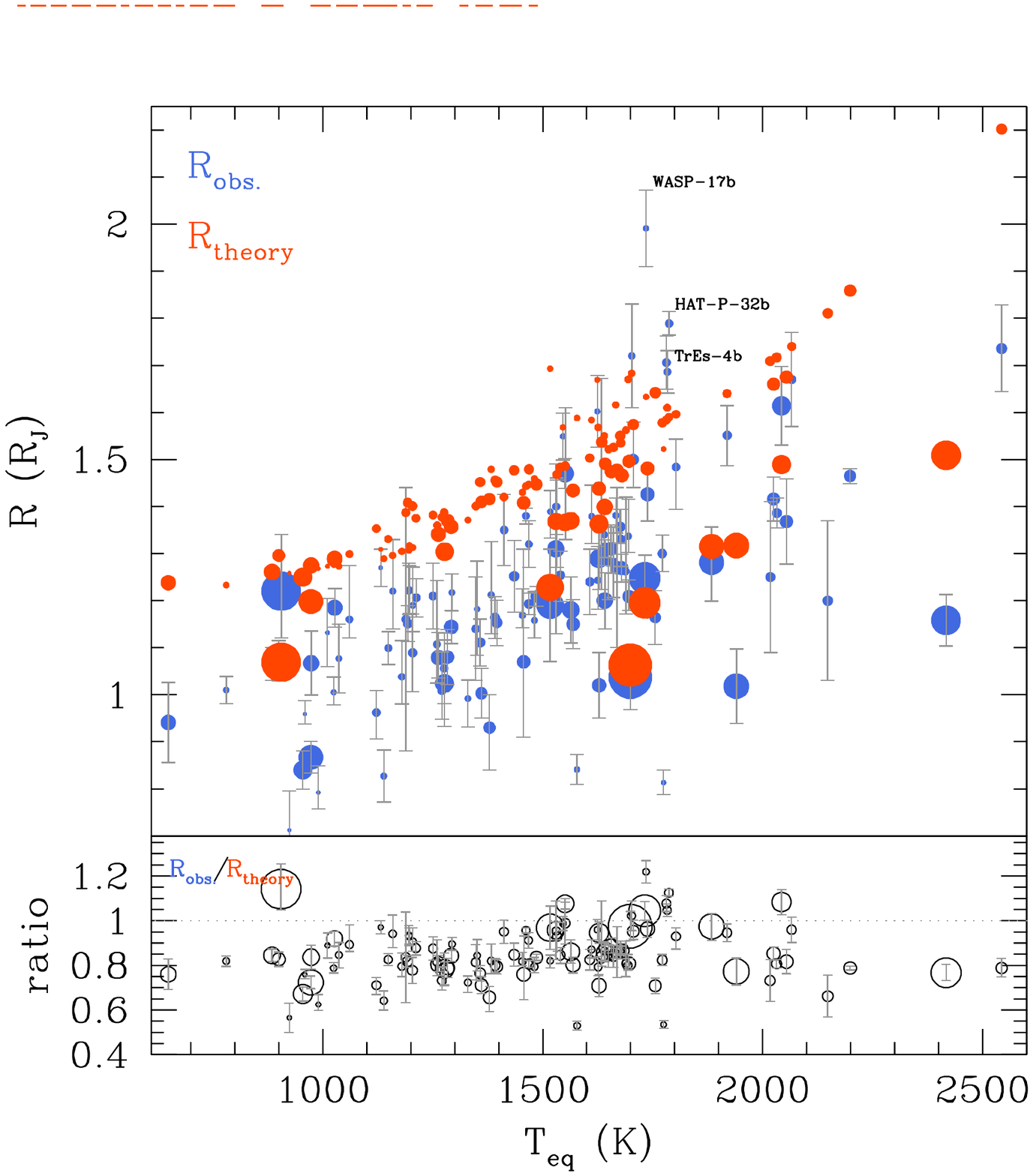}
\caption{Expected radii at 1 Gyr of age for the list of transiting
  planets as compiled by \citet{southworth}, plotted against their
  black-body equilibrium temperature.  The sizes of the symbols
  correspond to planet masses (ranging between $0.2 M_J$ and $30
  M_J$).  We have adopted $\epsilon=3\%$ and $z_{\rm wind} = 2.5\times
  10^8\cm$.
  Our predicted radius excess, $\Delta R = R-R_J \propto T_{\rm
    eq}^{1.2}$ for $1M_J$ planet, and $\Delta R \propto 1/\sqrt{M}$
  for planets with $T_{\rm eq} \approx 1500\K$.  In contrast to
  \citetalias{Batygin-ohmic}, our calculations of Ohmic heating does not
  lead to runaway expansion, even for low mass planets. The lower
  panel plots the ratio between the observed and theoretical
  radii. Most planets have radii below the theoretical expectations,
  with the exception of WASP-17b and HAT-P-32b that have measured
  radii more than $2\sigma$ above the theoretical values. }
\label{fig:ohmic-radius}
\end{center}
\end{figure*}

To study the contraction of a planet under Ohmic heating, we construct
a sequence of planet models that are in hydrodynamical and thermal
equilibrium. We include the effects of stellar insolation and Ohmic
heating; Ohmic heating is calculated self-consistently with an assumed
efficiency $\epsilon$ and wind depth $z_{wind}$. {\y In this section,
  we are interested in how Ohmic heating stalls the cooling
  contraction. So planets are assumed to be losing entropy and
  contracting in time. In \S \ref{sec:reinflation}, we adopt the MESA
  code \citep{mesa} to study the problem of re-inflating planets after
  they have contracted.}

A few details on the model building.  We interpolate the tables of
equation of state \citep{SCvH} and opacity \citep{Ferguson,
  Allard,Potekhin} relevant for planetary mass objects.  The center
boundary condition is a dense solid core of size $100\km$, a density
of $13\g/\cm^3$ and hence a small core mass of $\sim 5\times 10^{22}
\g$. The outer boundary condition is that at the photosphere
$p=(2/3)g/\kappa$ and $T = T_{\rm eq} = (L_*/16 \pi \sigma_B a^2
)^{1/4}$. We calculate the electron fraction for a solar composition
gas \citep{Anders}, taking into account the thermal and pressure
ionization of hydrogen, helium and three metals (Na, K and Al, these
are the elements with the lowest first ionization potential {\y as
  well as high cosmic abundances}).  Metal contributions dominate near
the surface while hydrogen is important in deeper regions.  Electron
conductivity is calculated using the electron-neutral collision
cross-section from \citet{Draine}.
The depth of the wind layer is typically set to be $z_{\rm wind} =
2.5\times 10^8 \cm$ (near pressure $p \sim 10$ bar),\footnote{
  In Appendix \ref{sec:force}, we address to what extent the Lorentz
  force can accelerate fluid below the wind layer. We show that the
  resulting fluid velocities are negligible, and hence the wind
  profile adopted here is self-consistent.} the surface dipole field
strength is $5$ Gauss.  The wind velocity is adjusted so that the
total Ohmic heating efficiency $\epsilon$ reaches the preset value,
typically $3\%$. The electric current is solved as a boundary value
problem (see Appendix \ref{sec:radial}), with $J_r = 0$ at the surface
and the electric potential $\Phi \propto r^2$ near the center.  Heat
is carried out by radiative diffusion, or convection where the
temperature gradient is super-adiabatic.  The planet model is iterated
until the profile of Ohmic heating and the thermal structure are
mutually compatible.

For each model, we must assume an intrinsic cooling luminosity $L_{\rm
  cool}$. This is the net energy that is lost from the planet
interior. The planet's internal entropy, as well as its size, are
related to the total luminosity flowing through $z_{\rm tr}$.  We
align models with different values of internal entropy into a time
sequence, using the relation of $\int T{{dS}\over{dt}} d^3x = L_{\rm
  cool}$. The Ohmic heating does not contribute to the right-hand-side
as it is an external energy source that hardly affects the bulk
entropy of the planet.

We have also compared the results from our equilibrium models with
those using MESA which integrates the temporal thermal evolution. We
confirm that when we use the same Ohmic heating profile in MESA as in
our equilibrium models, we obtain very similar results.


 Fig. \ref{fig:ohmic-noohmic-special} shows results for the planet
 TrEs4-b. The cooling curves in three different scenarios are
 plotted. If TrEs4-b was born and evolved in isolation, it would have
 cooled to its current measured radii ($1.7 R_J$) in less than $10$
 Myr; being irradiated by its host star at its current orbit raises
 this figure to $\sim 10^8$ yr; and having Ohmic heating on top of the
 irradiation prolongs it further to $\sim 10^9$ yr. Ohmic heating can
 be effective in stalling contraction.

 The radial profile of Ohmic heating is displayed in
 Fig. \ref{fig:ohmic-special-compare-3}, when TrEs-4b is at $1.6R_J$.
 Less than a percent of the total Ohmic heating is deposited below
 $z_{\rm tr}$ yet this amount is sufficient for suspending the
 contraction. Fig. \ref{fig:ohmic-noohmic-special} also demonstrates
 that when the wind layer is much shallower, the $\epsilon$ required
 is much larger (eq. \ref{eq:epscon}).

 {\y Lastly, the radial profiles of electrical conductivity and gas
   temperature are displayed in Fig. \ref{fig:ohmic-sigma-profile}.
   Compared to the models shown in Fig. 2 of \citetalias{Batygin-ohmic},
   this planet model has a larger radius, a higher internal entropy,
   and a shallower $z_{\rm tr}$. In particular, the intense Ohmic
   heating near the surface causes the region above $\sim 10$ bar to
   become convective. A similar model without Ohmic heating will have
   an isothermal atmosphere (Fig. \ref{fig:noohmic-special-T}). The
   surface convection zone can significantly alter the local
   hydrodynamics.}

 {\y The electrical conductivity we obtain differs from that shown in
   \citetalias{Batygin-ohmic}, but compares well against that in
   \citet{huang}, a paper that appeared after the submission of this
   work.  Given our assumption of constant total heating efficiency,
   the overall magnitude of $\sigma$ does not affect the results. But
   the radial profile (especially that near the surface) will. This
   may partly explain (see \S \ref{sec:caveats}) the difference
   between results here and those in \citetalias{Batygin-ohmic}.}

 We proceed to calculate the expected radii at 1 Gyr for the list of
 transiting planets, as compiled by \citet{southworth} (see
 http://www.astro.keele.ac.uk/jkt/tepcat), who has updated some of the
 values from previous publications through his homogeneous analysis.
These are shown in Fig. \ref{fig:ohmic-radius} for which we assume
that both insolation and Ohmic heating have acted on each planet since
birth.  We find that the expected radius excess can be roughly fitted
by the following expression
  \begin{equation} {{R - R_J}\over{R_J}} = {{\Delta R}\over {R_J}}
    \approx 0.5 \left({{T_{\rm eq}}\over{1500\K}}\right)^{1.2}\, ,
\label{eq:scaling1}
\end{equation}
for planets at $\sim 1 M_J$. This results from the increasing
conductivity and the larger irradiation luminosity when $T_{\rm eq}$
is raised. {\y This scaling is obtained for a constant heating
  efficiency of $3\%$ (see \S \ref{sec:caveats} for discussions). This
  scaling is compatible with the $\Delta R \propto T_{\rm eq}^{1.4\pm
    0.6}$ scaling obtained from the observed sample by
  \citet{Laughlin11}.}
For different planet masses, we find the excess decreases with mass:
$T_{\rm eq} \approx 1500\K$, the excess scales approximately as
$(R-R_J)/R_J \sim 0.5 (M/M_J)^{-1/2}$. More massive planets have
smaller pressure scale heights for the same entropy.

The theoretical radii exceed those of almost all known hot jupiters,
with the exceptions of a handful of planets at $T_{\rm eq} \sim 1700
\K$. Among these, WASP-17b \citep{wasp17b} and HAT-P-32b
\citep{Hartman} are larger than the theoretical values by more than
$2\sigma$.  In fact, these two planets are so large their interiors
are almost fully adiabatic.  For planets with such a high entropy,
$z_{\rm tr}$ is almost at the photosphere, and hence an efficiency of
Ohmic heating $\epsilon \geq 1$ (eq.  \ref{eq:epscon}) is required to
stop the loss of entropy.

\section{Can a Cold Planet be Re-inflated?}
\label{sec:reinflation}

\begin{figure}
\begin{center}
\includegraphics[width=0.50\textwidth,angle=0,
trim=20 120 35 150,clip=true]{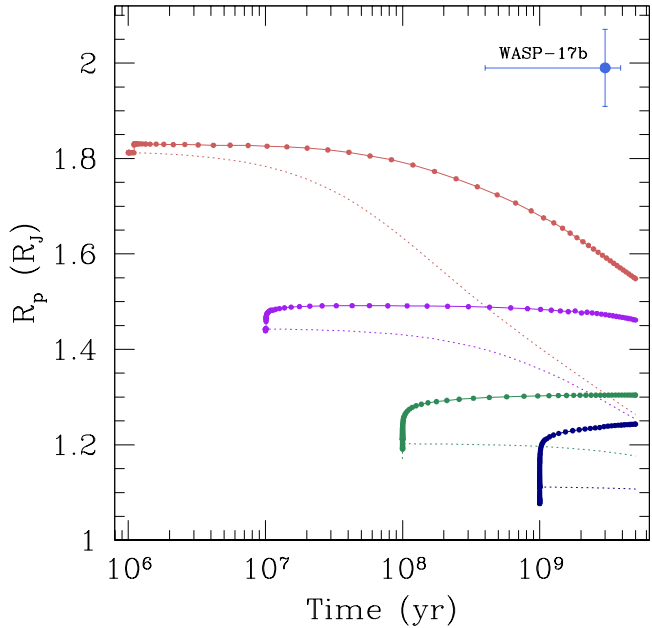}
\caption{The re-inflation of WASP-17b (the data point at the top
  right) by Ohmic heating ($\epsilon=3\%$) and stellar insolation
  ($T_{\rm eq} = 1770 \K$). Different tracks of radius evolution
  (solid curves with points) correspond to initial models that have
  cooled in isolation for $10^6, 10^7, 10^8$ and $10^9$ yrs,
  respectively (from top to bottom). Once the core entropy has been
  reduced to a low level, heating in the atmosphere only has a limited
  effect on the radius. The dotted curves show the same evolutionary
  models with stellar irradiation only.}
\label{fig:radiusexpansion}
\end{center}
\end{figure}

\begin{figure}
\begin{center}
\includegraphics[width=0.50\textwidth,angle=0,
trim=20 120 35 150,clip=true]{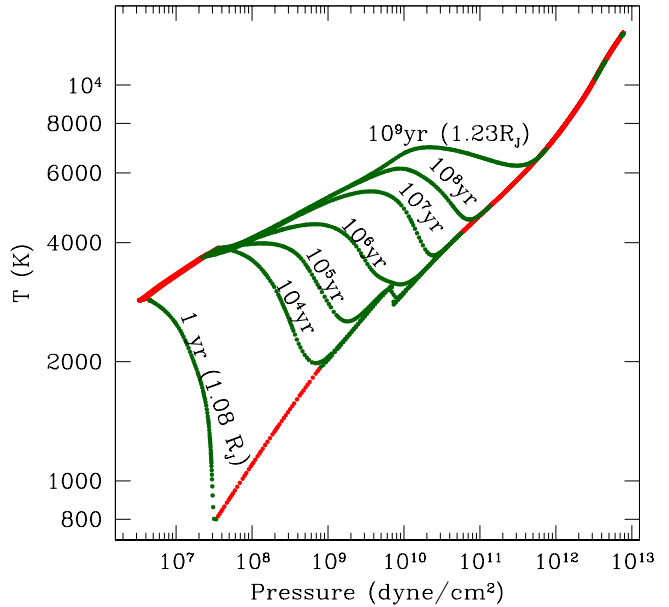}
\caption{Evolution of the temperature structure in a WASP-17b
    like planet, under Ohmic heating ($3\%$) and stellar irradiation,
    after the planet has cooled in isolation for 1 Gyrs (the lowest
    trajectory in Fig. \ref{fig:radiusexpansion}). The outer most
    point lies at an optical depth of $100$. The individual curves are
    snapshots taken at different epochs after the heating has
    commenced. The planet is eventually {\y mildly} 
re-inflated to $1.25 R_J$ after    5 Gyrs.
%
  With time, a thicker and thicker surface layer of the planet
    is lifted from the original adiabat to a higher adiabat. Ohmic
    heating below $p\approx 10^7 \dyne/cm^2$ is typically so small
    that it does not contribute much to the temperature rise
    there. Instead, heat diffuses downward gradually with the help of
    a temperature inversion.
    Red color indicates convective regions and green radiative
    regions.{\y The wind zone is convective.}}
\label{fig:reinflation}
\end{center}
\end{figure}

We turn to consider the scenario where the planet has contracted
significantly before it is subjected to intense stellar irradiation
and Ohmic heating. This is possible in a variety of migration theories
\citep[e.g.][]{WuMurray,fabtremaine,WuLithwick,Nagasawa} where the hot
jupiters arrive at their current locations a few million to a few
billion years after their formation.

We focus on WASP-17b \citep[$0.49 M_J, T_{\rm eq} =
1770\K$,][]{andersonage,wasp17b}, one of the lowest mass hot jupiters
that is also heavily irradiated. These two factors should combine to
produce the most pronounced re-inflation.\footnote{We note that the
  observed radius of WASP-17b is too big to explain even if Ohmic
  heating commences at formation. We use its extreme properties to
  illustrate the inefficacy of Ohmic re-inflation.}  Since this is a
non-equilibrium phenomenon, we employ the MESA package \citep{mesa},
which has the option of a user-supplied routine for additional heating
sources.  The heating source we adopt has an energy generation rate
per unit mass that scales quadratically with pressure above $z_{\rm
  wind}$, dropping by a factor of $(R/z_{\rm wind})^2$ at $z_{\rm
  wind}$, and scales as $p^{-1.3}$ below the wind zone until
$p=10^{10} \dyne/\cm^2$, below which the high conductivity leads to a
diminishingly small heating that scales as $p^{-4}$.  Such a profile
(the integrated heating is shown in
Fig. \ref{fig:ohmic-special-compare-3}) closely approximates the
actual Ohmic heating profile we calculate for WASP-17b
  and agrees with our analytical model in Appendix \ref{sec:radial}. 
  The results on radius expansion are shown in
  Fig. \ref{fig:radiusexpansion} where planet models are exposed to
  both Ohmic heating ($3\%$) and surface irradiation (with $T = T_{\rm
    eq}$), after having cooled in isolation for a period ranging from
  $10^6$ to $10^9$ yrs.

  The MESA calculations show that re-inflation is ineffective: the
  planet can at best be re-inflated by $\sim 10\%$ if it has already
  cooled (Fig. \ref{fig:radiusexpansion}). In
  Fig. \ref{fig:reinflation}, we illustrate in detail why this is so.
  Ohmic heating is intrinsically shallow and most of the energy is
  deposited at $p < 10$ bar $=10^7$ dyn/cm$^2$, whereupon most of it
  is quickly transported outward by the {\y newly generated} surface
  convection zone. The interior of the planet experiences the heating
  only when a temperature inversion is built up and heat diffuses
  gradually inward.\footnote{\y It is often claimed that as long as
    heat is deposited somewhere in the convection zone, it will be
    mixed uniformly. This is not true. Heat can not be transported
    against the temperature gradient, except with the help of external
    forcing.} By the time heat has diffused downward to $p = 10^{11}
  \dyne/\cm^2$, the temperature inversion is minimal and the radiative
  diffusion has nearly ground to a halt.  As a result, only layers
  shallower than this experience appreciable heating.  Since the
  pressure scale height is
$\sim c_s^2/g \sim 4\% R_J (T/7000\K)^{1/2}$,
the planet can at most expand a few times this value.  {\y This result
  has significance for the survival of hot jupiters.}

{\y In contrast to our result here, \citetalias{Batygin-ohmic} find
  that low mass planets can be reinflated by Ohmic heating to become
  Roche Lobe overflowing. In fact, \citetalias{Batygin-ohmic}
  typically find a more drastic effect of Ohmic heating than that 
  obtained here: our TrEs-4b model reaches an equilibrium radius of
  $1.6 R_J$ (Fig. \ref{fig:ohmic-noohmic-special}), while a similar
  model in \citetalias{Batygin-ohmic} yields $1.9 R_J$
  \citepalias[Fig. 5 in][]{Batygin-ohmic}.  Detailed differences in,
  e.g., radial profiles of conductivity and wind, treatment of
  radiative atmosphere, between our models may go some way in
  explaining this difference. However, re-inflation by of order unity
  requires raising temperature in the planetary center (bottom scale
  height). We find that radiative diffusion could not accomplish this
  (Fig. \ref{fig:reinflation}).}


Fig. \ref{fig:reinflation} also demonstrates that heating a cold
planet from the outside produces a {\y much more pronounced radiative
  region (compare with Fig. \ref{fig:ohmic-sigma-profile}).  This may
  have consequences for tidal dissipation, elemental diffusion and
  magnetic dynamo in the planet.}

In conclusion, we find that it is difficult to re-inflate by a
substantial amount a planet that has contracted.

\section{Discussion on Uncertainties}
\label{sec:caveats}

{\y We clarify two issues here.}
{\y Our assumption of a fixed Ohmic efficiency is ad hoc.  It allows
  us to focus on the thermodynamics of Ohmic heating, but obscures the
  atmospheric physics that determines the actual wind speed, as well
  as any dependence on planet temperature or magnetic field strength.
  \citet{menou11} and \citetalias{Batygin-ohmic} have argued that, in
  the weather layer ($\sim 0.1 bar$) where stellar insolation produces
  a large horizontal pressure gradient, the wind is driven by the
  thermal gradient but dragged by the Lorentz force. Assuming the two
  forces balance each other,
  \citet{menou11} obtains an equilibrium wind speed and an Ohmic
  heating efficiency (in the weather layer) that peaks at a few
  percent when the surface temperature is $\approx 1500\K$ ($B\sim
  10\, {\rm G}$). However, this argument may be incomplete. The
  biggest Lorentz drag is applied not at the weather layer, but deeper
  down, at the bottom of the wind zone where conductivity is the
  highest \citep[see, e.g.][]{Liu}. So the actual wind speed depends
  on how deep the wind zone extends to. This in turns depends on how
  well the weather layer couples dynamically to the deeper atmosphere,
  a process currently not well understood
  \citep{2010ApJ...719.1421P,Rauscher,LianShowman,SchneiderLiu}. An
  added complication is that Ohmic heating itself can modify the local
  temperature gradient and stratification (see below). Further study
  is necessary.}

  An additional question is whether heat deposited above $z_{\rm tr}$
  can affect the radius evolution. This is the majority of Ohmic
  heating and can be characterized by GCM simulations. We argue (\S
  \ref{sec:thermal}) that this fraction of heat cannot directly
  supplant the internal entropy loss and halt the contraction. But it
  raises the local temperature gradient
  (Fig. \ref{fig:ohmic-sigma-profile})\footnote{In more extreme cases
    (Fig. \ref{fig:reinflation}), temperature gradient near the
    surface is even inverted.}  and exerts an indirect effect on the
  radius evolution, as is observed in our numerical results. At a
  given interior adiabat, an atmosphere with such a superficial
  heating has a deeper $z_{\rm tr}$ than an atmosphere that is only
  irradiated and isothermal.  Deeper $z_{\rm tr}$ implies that the
  amount of energy necessary to suspend internal cooling, the one that
  is deposited below $z_{\rm tr}$), is reduced (see \S
  \ref{sec:thermal}). The rising temperature in the wind zone brings
  about a greater conductivity, a larger magnetic drag, and likely, a
  smaller wind speed. This could lead to saturation of Ohmic heating.

}

\section{Conclusion}

We study the radius evolution of hot jupiters under Ohmic heating and
stellar insolation.  We consider initial conditions in which either
(a) the planet starts with an high entropy, or (b) it has previously
cooled and contracted. In the former case, we find that planets can be
sustained at large radii for billions of years
(Fig. \ref{fig:ohmic-radius}). For our example case of TrEs-4b, the
planet may remain in a state of perpetual youth, at a cooling age of
$10$ Myr (Fig. \ref{fig:ohmic-noohmic-special}). Less massive planets
can be suspended at larger radii than the more massive ones in the
same radiation environment.  We place a constraint on the Ohmic
heating if it is to suspend the cooling contraction: for an efficiency
of a few percent, the wind layer needs to extend to a depth of order
$10$ bar; a shallower layer would require a higher heating efficiency
(eq. \ref{eq:epscon}).

We also consider Ohmic heating on planet models that have previously
contracted.  Ohmic heating, being very superficial, has a limited
capacity for re-inflating these planets. We illustrate this using
WASP-17b, a low mass planet that has one of the highest irradiation
fluxes, yet still cannot be significantly inflated after cooling
(Fig. \ref{fig:radiusexpansion}).  As such, Ohmic heating is best
termed 'Ohmic suspension', rather than 'Ohmic inflation'.

The interior structure of the planet {\y is affected by Ohmic
  heating}: the planet has an adiabatic interior and a largely
isothermal atmosphere in the case {\y of no Ohmic heating}, but {\y
  the wind zone tends to become convective when Ohmic heating is
  applied.}  This may impact on the physics of {\y global
  circulation, magnetic dynamo,} tidal dissipation and element
settling.

While a few planets have observed radii comparable to or even larger
than predictions using our fiducial Ohmic heating parameters (WASP-17b
and HAT-P-32b, in particular), the vast majority fall below these
predictions.
In fact, many are consistent with cooling without any Ohmic heating,
and some are too small for pure gaseous spheres. There could be a
number of explanations for these. They could have lower Ohmic heating
efficiencies than stipulated here, or they could have massive solid
cores and/or heavy-metal enriched envelopes \citep[see,
e.g.][]{Baraffe2}, or, alternatively, { they may have} a different
migration history, as we now describe.  Consider the scenario where
the planet is migrated to its current location well after
formation. Before the migration, it contracts and loses entropy in
isolation. If during the migration there is no significant entropy
injection into the deep interior of the planet, the planet will then
follow one of the trajectories in Fig. \ref{fig:radiusexpansion} and
end up having a different radius depending on its time of
migration. If this scenario is correct, it would mean that the present
day planet radii retain the memory of their past dynamical history.

{There is a caveat to our above scenario. During the migration
  process, there is likely a large amount of tidal energy deposited
  inside the planet and this may have rejuvenated the planet.
  However, there are substantial theoretical uncertainties regarding
  the mechanism and location of tidal heating. We hope to address
  tidal rejuvenation in an upcoming publication.}

The super-sizes of WASP-17b and HAT-P-32b are surprising: if Ohmic
heating is responsible for their sizes, it must be both very efficient
and remarkably continuous. For instance, {if the dynamo field weakens
  for, say, $10^8$ years, the internal entropy would be mostly drained
  away and the planet would have contracted.}

{\y We thank the referee, Kristen Menou, for suggestions that improved
  the clarity of this paper.}  Y.W acknowledges useful conversations
with {P. Goldreich}, T. Guillot, K. Batygin, P. Arras, and especially
thanks B. Paxton and other MESA creators for this wonderful package.
Y.L. acknowledges support from NSF grant AST-1109776.

\bibliographystyle{apj}
\bibliography{ref}


\begin{appendix}

\section{Appendix A: The radial profile of Ohmic Heating}
\label{sec:radial}

Here we describe our model for Ohmic heating.  We assume a fixed wind
profile near the surface of the planet, and a dipolar magnetic field
threading the planet.  We ignore here the effect of the Lorentz force
on the fluid velocity, but return to discuss it in Appendix
\ref{sec:force}.  The electric current is obtained by solving \beqn
\bld{J} &=& \sigma \left( -{\bf \nabla}\Phi + {{{\bf v_{\rm
          wind}}\times {\bf B_{\rm dipole}}}\over c}\right)
\label{eq:jeqny}
\\
\bld{\nabla\cdot J}&=&0 \ , \eeqn subject to the boundary conditions
that $J_r=0$ at the planet's surface ($r=R$) and center ($r=0$).  We
have used that the steady state electric field is the gradient of a
potential, $\bld{E}=-\bld{\nabla}\Phi$.  The dipolar field is \beqn
\bld{B_{\rm dipole}}=-\bnabla\left(\bM\cdot\br\over
  r^3\right)=M\frac{(2\cos\theta\ber+\sin\theta\bet)}{r^3}\, ,
\label{eq:By}
\eeqn
in spherical co-ordinates, and we assume that the wind has a constant
angular velocity $\omega$ in a zone that extends to a depth $\delta
R$, i.e.  \beqn
\bld{v_{\rm wind}}= \omega r \sin\theta \bld{e}_\phi \ \ \ \ {\rm for\ }  1-\delta < r/R <1 \\
\eeqn and is zero for $r/R<1-\delta$.  The above equations combine to
yield
\beqn \bld{J}=-\sigma \bld{\nabla}\left( \Phi+\frac{2\omega
    M}{3cr}\left(P_2-P_0\right) \right) \ \ \ \ {\rm for\ } 1-\delta <
r/R <1 \eeqn
and $\bld{J}=-\sigma\bld{\nabla}\Phi$ below the wind zone.  We discard
the uninteresting term $P_0$ and assume all variables follow a $P_2$
dependence, adopting the
form
$\Phi= \Phi_2(r)P_2$ for the potential.
In the body of the paper, we numerically solve the ODE for $\Phi_2$
that results from setting $\bld{\nabla\cdot J}=0$ to obtain the Ohmic
heating rate, which is the volume-integrated $J^2/\sigma$.

In the remainder of this Appendix, we solve analytically a simple
example that illustrates two important features of Ohmic heating:
first, the rapid drop in the heating profile by the factor $\sim
\delta^2$ just below the wind, and second, the inward decay in the
heating profile in the interior (see numerical results in Fig.
\ref{fig:ohmic-special-compare-3}).  We take $\sigma$ to be {\y a}
piecewise {\y function of radius}, with $\sigma=\sigma_{shallow}$
through the wind zone to a depth $\Delta R$, where $\Delta>\delta$;
and $\sigma=\sigma_{deep}$ below that.  There are then three shells of
interest.  In the outer one ($ 1-\delta < r/R <1$), there is a wind
and $\sigma=\sigma_{shallow}$; in the middle one ($ 1-\Delta < r/R
<1-\delta$), there is no wind and $\sigma=\sigma_{shallow}$; and in
the inner one ($r/R<1-\Delta$) there is no wind and
$\sigma=\sigma_{deep}$.  The ODE for $\Phi_2$ is easily solved
analytically in each shell, and between the shells one requires that
$\Phi$ and $J_r$ are continuous.  Omitting the algebra, we find that
in the limit $\delta<\Delta\ll 1$ and $\sigma_{shallow}\ll
\sigma_{deep}$ the current in each of the three shells is given by
\beqn {\rm Outer\ shell \ (wind\ zone)}:&\ \ \ J_r\approx -6 \left(
  1-{r\over R} \right) \cdot J_0 P_2 \ , \ \ \ \ \ J_\theta\approx J_0
{dP_2/ d\theta}\, ,
\label{eq:outer} \\
{\rm Middle\ shell}:&\ \ \ J_r\approx -6\delta \cdot J_0 P_2 \ , \ \ \
\ \
J_\theta\approx 6\delta(1-{r\over R}-\Delta)  \cdot J_0 {dP_2/d\theta}\, ,
\label{eq:middle} \\
{\rm Inner\ shell}:&\ \ \ J_r\approx -6{r\over R}\delta \cdot J_0 P_2 \
, \ \ \ \ \ J_\theta\approx -3{r\over R}\delta \cdot J_0 dP_2/d\theta
\eeqn where the constant \beqn J_0 \equiv {2\omega
  M\sigma_{shallow}\over 3 c R^2}\, , \label{eq:inner}
\eeqn 
Hence the total current in the
wind zone is dominated by $J_\theta$ which is $\sim 1/\delta$ greater
than the radial current $J_r$. Below the wind zone, $J_r$ is
continuous while $J_\theta$ drops discontinuously to less than
$J_r$. This causes a large drop in the rate of Ohmic heating
($J^2/\sigma$) below the wind zone, of order $\delta^2 = (z_{\rm
  wind}/R)^2$. Moreover, the higher conductivity in the core ensures
that there is little Ohmic dissipation there.  Substituting values of
$\delta=0.001$, and $\Delta=0.005$ and
$\sigma_{deep}/\sigma_{shallow}=10^7$, the fractions of the total
ohmic dissipation taking place in the outer, middle, and inner shells
are $0.999855, 0.000144851, 9.04343\times10^{-10}$, respectively
We argue in the text that only heating in the middle shell can
contribute to halting the planet's contraction.


\section{Appendix B: Dynamical Effects Below the Wind Zone}
\label{sec:force}

While wind is forced in the outer shell, there is no external forcing
in the inner shells.  A simple estimate for the Lorentz force (${\bf
  J}\times {\bf B}/c$) shows that the inner layers could be spun up in
a short timescale. We investigate the dynamical response of the planet
to the surface battery by looking for steady-state solution to the
following equations,
\begin{eqnarray}
\rho {{\partial{\bf v}}\over{\partial t}} & = & {{{\bf J}\times {\bf B}}\over c} \label{eq:eq1}\\
{1\over c} {{\partial{\bf b}}\over{\partial t}} & = & - {\bf \nabla}\times
{\bf E}\, . \label{eq:eq2}
\end{eqnarray}
Here, ${\bf b}$ denotes perturbed field line, while ${\bf B}$ the
background field. 
At steady state, ${\bf J} = {\bf \nabla}\times {\bf b}$.

We will study a cylindrical analog to the planet where the cylindrical
$z$ direction approximates that of the spherical $\theta$ direction,
and cylindrical radius that of the spherical radius. The dipole field
is now ${\bf B} = B_r {\bf e_r}$ with $B_r = B_0 R/r$. We include only
the components ${\bf b} = b_\phi {\bf e_\phi}$ and ${\bf j} = j_z {\bf
  e_z}$. We further assume a harmonic $z$ dependence $\propto e^{i k_z
  z}$.

In the wind zone, an external forcing, ${\bf F} = F{\bf e_\phi}$ maintains
the current. The time-{\y in}dependent equations are now,
\begin{eqnarray}
 {{B_r}\over c} {1\over r} {\partial \over{\partial r}} \left(r b_\phi\right)
 + F & = & 0\nonumber \\
- {{k_z^2}\over{\sigma}}b_{\phi} + {\partial
    \over{\partial r}} \left( {1\over {\sigma r}} {\partial
      \over{\partial r}} \left(r b_\phi\right)\right) - {{B_r}\over c} 
{{\partial v_{\phi}}\over{\partial r}}
& = & 0\, .
\label{eq:bphi_wind}
\end{eqnarray}
The boundary condition is that the radial current vanishes at $r=R$,
or $b_\phi = 0$ at $R$.  Taking $r\approx R$, and a constant forcing
$F$, we find the solution to be
\begin{eqnarray}
b_\phi & = &  {{F c R}\over{B_0}} (1-r/R)\nonumber \\
v_\phi & \approx & v_{\rm wind} =  {\rm const.}
\label{eq:windsol}
\end{eqnarray}
since we take the wind zone be of thickness $\delta \ll 1 \sim 1/(Rk_z)$.

At the bottom of the wind zone, $b_\phi$ behaves continuously (radial
current has to be continuous), but a jump in the tangential velocity
($\Delta v$) is necessary to ensure that the tangential electric field
is continuous. The velocity there is
\begin{equation}
v_\phi (r = R-\delta)= v_{\rm wind}- \Delta v = v_{\rm wind} - {{F c^2}\over{B_r^2 \sigma}}.
\end{equation}
Further below, both the velocity and $b_\phi$ decay inward as,
\begin{eqnarray}
b_\phi & = & {{F c R \delta }\over{B_0 r}} e^{i k_z z}\nonumber \\
{{\partial v_\phi}\over{\partial r}} & = & -  {{F c^2 k_z^2 
 \delta  }\over{B_0^2  \sigma}} e^{i k_z z}.
\end{eqnarray}
So the region of lowest $\sigma$ is also where $v_\phi$ decreases most
steeply. Compared to $\Delta v$, the total shear across the no-forcing
zone is much smaller,
\begin{equation} {v_\phi (r = R-\delta) - v_\phi(r=0)} \sim k_z^2
  \delta^2 \, \Delta v \ll \Delta v\, .
\end{equation}
This result indicates that, at steady state, the interior is able to
maintain a largely uniform rotation with $v_\phi \approx v_\phi
(r=R-\delta)$. Contrary to our initial expectation, the interior can
rotate at a different rate than that in the wind zone, despite the
large Lorentz torque. {\y It is not spun up.}

A few words on the energetics. Stellar insolation drives the wind
which suffers both turbulent dissipation and Ohmic loss. We take the
the timescale for turbulent dissipation of the wind to be the wind
travelling time across the planet, $R/v_{\rm wind}$
\begin{equation}
\left({{dE}\over{dt}}\right)_{\rm turb} \approx
2 \pi r \delta {{\rho v_{\rm wind}^3}\over R}\, ,
\end{equation}
and the rate for Ohmic loss is
\begin{equation}
\left({{dE}\over{dt}}\right)_{\rm ohmic} \approx
2 \pi r \delta F v_{\rm wind} 
= 2 \pi r \delta v_{\rm wind}^2 {{B_r^2 \sigma}\over {c^2}}\, .
\end{equation}
We find the ratio between the two rates to be 
\begin{equation}
{{\left({{dE}/{dt}}\right)_{\rm turb} }\over
{\left({{dE}/{dt}}\right)_{\rm ohmic} }} \approx
{{\rho v_{\rm wind} c^2}\over{R B_r^2 \sigma}}\, .
\end{equation}
Adopting parameters $\rho = 10^{-4}\g/\cm^3$, $\sigma=10^{10}\s^{-1}$,
$B_r = 5$G and $v_{\rm wind} = 10^5 \cm/\s$, this ratio is $\approx
4$, or Ohmic loss is non-negligible. This suggest that the Ohmic
efficiency may be higher than the canonical $3\%$ we adopt in the
text.

\end{appendix}

\end{document}